\documentclass[a4paper,10pt]{article}

\usepackage[english]{babel}
\usepackage[utf8]{inputenc}
\usepackage[T1]{fontenc}
\usepackage[babel=true]{csquotes}

\usepackage{amsfonts,amssymb,amsmath,amsthm}
\allowdisplaybreaks

\DeclareMathOperator{\diag}{diag}
\usepackage[usenames,dvipsnames]{pstricks}
\usepackage{epsfig}
\usepackage{pst-grad} 
\usepackage{pst-plot} 
\usepackage{multirow}
\usepackage{graphicx}
\usepackage[footnotesize]{caption}
\usepackage{color,xspace}
\usepackage{epstopdf}
\graphicspath{{figures_vba/}}

\newcommand{\poubelle}[1]{}

\usepackage[top=1.5cm, left=1.5cm, right=1.5cm, bottom=1.5cm]{geometry}

\usepackage{alphabet1}
\usepackage{alphabet2}
\usepackage{macros_gpi}

\title{Variational Bayesian Approach and Gauss-Markov-Potts prior model}

\author{Camille Chapdelaine$^{1,2}$
\\
  \footnotesize
  $^1$ Laboratoire des signaux et syst\`emes, CNRS, CentraleSup\'elec-Universit\'e Paris Saclay, Gif-sur-Yvette, France\\
  \footnotesize
  $^2$ SAFRAN SA, Safran Tech, P\^ole Technologie du Signal et de l'Information, Magny-Les-Hameaux, France\\   
  } 
\date{June 2018}

\begin{document}

\maketitle

\section{Introduction}
\label{Introduction}

In many inverse problems such as 3D X-ray Computed Tomography (CT) \cite{fessler2000statistical}, the estimation of an unknown quantity, such as a volume or an image, can be greatly enhanced, compared to maximum-likelihood techniques \cite{feldkamp1984practical, jackson1991selection, gordon1970algebraic, gilbert1972iterative, andersen1984simultaneous}, by incorporating a prior model on the quantity to reconstruct.

This prior model is often defined in terms of sparsity on the unknown in some domain, for instance sparsity of one of its derivative \cite{sidky2012convex, storath2015joint}, of a wavelet transform \cite{wang2017x, notargiacomo2016sparse}, or of its representation in a learnt dictionary \cite{chun2017sparse, zheng2017union}. A more complex prior can be designed for multi-channel estimation such as reconstruction and segmentation thanks to Gauss-Markov-Potts prior model \cite{feron2005microwave, mohammad2008gauss, AyDj10, zhao2016joint, chapdelaine20163DBayesian}.

For very large inverse problems such as 3D X-ray CT, maximization a posteriori (MAP) techniques are often used due to the huge size of the data and the unknown \cite{kim2013accelerating, kim2015combining, mcgaffin2015alternating}. Nevertheless, MAP estimation does not enable to have quantify uncertainties on the retrieved reconstruction, which can be useful for post-reconstruction processes for instance in industry and medicine. In X-ray CT, a method has been proposed in \cite{fessler1996mean} to estimate exact uncertainties but can only be applied to few pixels of interest. More recently, in \cite{pereyra2017maximum}, an estimation of confidence regions for MAP estimator is detailed, but is difficult to apply for joint reconstruction and segmentation algorithms \cite{chapdelaine20163DBayesian}. Another way to tackle the problem of uncertainties estimation is to compute posterior mean (PM) for which the uncertainties are the variances of the posterior distribution. Because MCMC methods are not affordable for very large 3D problems, this paper presents an algorithm to jointly estimate the reconstruction and the uncertainties by computing PM thanks to variational Bayesian approach (VBA) \cite{SQ06, pereyra2016survey}. The prior model we consider for the unknowns is a Gauss-Markov-Potts prior which has been shown to give good results in many inverse problems \cite{feron2005microwave, bali2008bayesian, ayasso2010bayesian, zhao2016joint, chapdelaine20163DBayesian, giovannelli2017deconvolution}. After having detailed the used prior models, the algorithm based on VBA is detailed : it corresponds to an iterative computation of approximate distributions through the iterative updates of their parameters. The updating formulae are given in the last section. We also provide a method for initialization of the algorithm, as a method to fix each parameter. Perspectives are applications of this algorithm to large 3D problems such as 3D X-ray CT.  

\section{Prior models}
\label{sec:Modèles utilisés}

\subsection{Forward model}
\label{sec:Modèle pour les projections}

We consider a general forward model for linear inverse problems, accounting for uncertainties
\beq
\gb=\Hb\fb+\zetab
\label{eq:Modèle direct usuel}
\eeq
For instance, this forward model is used in 3D X-ray CT : $\fb$ is the volume to reconstruct, and is discretized in $N=N_x\times N_y\times N_z$ voxels. We denote by $M=$ the number of measurements, which is the size of $\gb$. In 3D X-ray CT, matrix $\Hb$, which is size $M\times N$, is called the projection operator or projector. Its adjoint $\Hb^T$ is called the backprojection operator or backprojector. Uncertainties $\zetab$ are modeled as Gaussian \cite{sauer1993local}
\beq
p(\zeta_i|\rho_{\zeta_i})=\Nc(\zeta_i|0,\rho_{\zeta_i}^{-1})
\eeq
A conjugate prior is assigned to inverse variances $\rhob_{\zeta}$ :
\beq
p(\rho_{\zeta_i}|\alpha_{\zeta_0},\beta_{\zeta_0})=\Gc(\rho_{\zeta_i}|\alpha_{\zeta_0},\beta_{\zeta_0}).
\eeq
$\Gc$ denotes Gamma distribution
\beq
\Gc(\rho_{\zeta_i}|\alpha_{\zeta_0},\beta_{\zeta_0})=\frac{\beta_{\zeta_0}^{\alpha_{\zeta_0}}}{\Gamma(\alpha_{\zeta_0})}\rho_{\zeta_{i}}^{\alpha_{\zeta_0}-1}
\expf{-\beta_{\zeta_0}\rho_{\zeta_{i}}}, \rho_{\zeta_i}>0, \forall i
\eeq
where $\Gamma$ is Euler's gamma function.

\subsection{Gauss-Markov-Potts prior model for the volume}
\label{sec:Modèle de Gauss-Markov-Potts pour le volume}

Gauss-Markov-Potts prior model introduces a dependance of $f_j$ on the material in which voxel $j$ is \cite{mohammad2008gauss, Ayasso10, giovannelli2017deconvolution}. Each voxel is assigned a label $z_j$ which is $z_j=k$ if voxel $j$ is in material $k$, $k \in \mathbb{N}$, $1\leq k\leq K$. $K$ is the number of materials and is supposed to be known. Given the material of voxel $j$, we have the following prior for $f_j$ : 
\beq
f_j \sim \Nc(m_{k},\rho_{k}^{-1}) \mbox{~~if~~} z_j=k.
\eeq
Means and inverse variances of the classes are unknown and are assigned a conjugate prior :
\beq
p(m_k|m_0,v_0)=\Nc(m_k|m_0,v_0)
\eeq
and 
\beq
p(\rho_k|\alpha_0,\beta_0)=\Gc(\rho_k|\alpha_0,\beta_0)
\eeq
where $m_0$, $v_0$, $\alpha_0$ et $\beta_0$ are fixed parameters.

\begin{figure}
 \begin{minipage}[htb]{0.24\linewidth}
  \centering
  \includegraphics[scale=3]{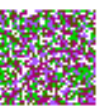}
  \caption*{$\gamma_0=0.5$}
 \end{minipage} \hfill
 \begin{minipage}[htb]{0.24\linewidth}
  \centering
  \includegraphics[scale=3]{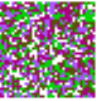}
  \caption*{$\gamma_0=0.7$}
 \end{minipage} \hfill
 \begin{minipage}[htb]{0.24\linewidth}
  \centering
  \includegraphics[scale=3]{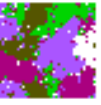}
  \caption*{$\gamma_0=0.8$}
 \end{minipage} \hfill
 \begin{minipage}[htb]{0.24\linewidth}
  \centering
  \includegraphics[scale=3]{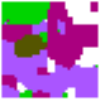}
  \caption*{$\gamma_0=1.6$}
 \end{minipage} 
 \caption{Potts fields $\zb$ for different values of $\gamma_0$}
 \label{fig:Variation de la compacité du champ de Potts en fonction de gamma0}
\end{figure}

A Potts prior is assigned to labels $\zb$ in order to promote compact regions in the volume \cite{feron2005microwave, AyDj10, chapdelaine20163DBayesian, giovannelli2017deconvolution}. Using Hammersley-Clifford theorem \cite{besag1974spatial}, this prior reads \cite{AyDj10, zhao2016joint, chapdelaine20163DBayesian} :
\beq 
p(\zb|\alphab,\gamma_0)\propto 
\expf{\sum_{j=1}^{N} \left(\sum_{k=1}^K \alpha_k\delta(z_j-k) + \gamma_0 \sum_{i\in\Vc(j)}\delta(z_j-z_i)\right)}
\label{eq:Champ général de Potts}
\eeq
where \cite{AyDj10}
\beq
\sum_{k=1}^{K} \expf{\alpha_k}=1.
\eeq
Parameter $\gamma_0$ is called Potts coefficient or granularity coefficient \cite{onsager1944crystal, huang1987statistical, giovannelli2010estimation, pereyra2013estimating, chapdelaine20163DBayesian}. It tunes the compacity of the classes, as shown in figure \ref{fig:Variation de la compacité du champ de Potts en fonction de gamma0}. Partition function for $\zb$ is
\beq
Z(\alphab,\gamma_0)=\sum_{\zb'\in\left\{1,\dots,K\right\}^N}\expf{\sum_{j=1}^{N} \left(\sum_{k=1}^K \alpha_k\delta(z_{j}'-k) + \gamma_0 \sum_{i\in\Vc(j)}\delta(z_{j}'-z_{i}')\right)}
\eeq
and is untractable \cite{morris1997fully, pereyra2013estimating}.

\section{Bayesian inference and variational Bayesian approach}
\label{sec:Description de l'approche bayésienne variationnelle}

Based on prior models $\Mc$ described in section \ref{sec:Modèles utilisés}, the joint posterior distribution of the unknowns 
\beq
\psib=(\fb,\rhob_{\zeta},\zb,\mb,\rhob)
\eeq
reads, according to Bayes'rule
\begin{align}
& p(\psib|\gb;\Mc)=p(\fb,\rhob_{\zeta},\zb,\mb,\rhob|\gb;\Mc) \nonumber \\
&=\frac{p(\gb;\psib|\Mc)}{p(\gb|\Mc)}=\frac{p(\gb;\fb,\rhob_{\zeta},\zb,\mb,\rhob|\Mc)}{p(\gb|\Mc)} \nonumber \\
&=\frac{p(\gb|\fb,\rhob_{\zeta})p(\fb|\zb,\mb,\rhob)p(\rhob_{\zeta}|\alpha_{\zeta_0},\beta_{\zeta_0})p(\zb|\alphab,\gamma_0)p(\mb|m_0,v_0)p(\rhob|\alpha_0,\beta_0)}{p(\gb|\Mc)}
\label{eq:posterior distribution}
\end{align}
where
\beq
p(\gb|\fb,\rhob_{\zeta})=(2\pi)^{-\frac{M}{2}}\det{\Vb_{\zeta}}^{-1/2}\expf{-\frac{1}{2}\|\gb-\Hb\fb\|_{\Vb_{\zeta}}^2},\\
\eeq
\beq
p(\fb|\zb,\mb,\rhob)=(2\pi)^{-\frac{N}{2}}\det{\Vb_{\zb}}^{-1/2}\expf{-\frac{1}{2}\|\fb-\mb_{\zb}\|_{\Vb_{\zb}}^2},\\
\eeq
\beq
p(\rhob_{\zeta}|\alpha_{\zeta_0},\beta_{\zeta_0})=\frac{\beta_{\zeta_0}^{\alpha_{\zeta_0}}}{\Gamma(\alpha_{\zeta_0})} 
\expf{\sum_{i=1}^M \left(((\alpha_{\zeta_0}-1)\ln \rho_{\zeta_{i}}-\beta_{\zeta_0}\rho_{\zeta_{i}}\right)},\\
\eeq
\beq
p(\zb|\alphab, \gamma_0)=\frac{1}{Z(\alphab,\gamma_0)} 
\expf{\sum_{j} \left(\sum_{k=1}^K \alpha_k \delta(z_j-k) + \gamma_0 \sum_{i\in\Vc(j)}\delta(z_j-z_i)\right)}
\eeq
\beq
p(\mb|m_{0}, v_{0})=(2\pi)^{-\frac{K}{2}}{v_{0}}^{-\frac{K}{2}}
\expf{-\frac{1}{2 v_{0}}\sum_{k=1}^{K} (m_{k}-m_{0})^2},\\
\eeq
\beq
p(\rhob|\alpha_0,\beta_0)=\frac{\beta_{0}^{\alpha_0}}{\Gamma(\alpha_0)} 
\expf{\sum_{k=1}^{K} \left((\alpha_0-1)\ln \rho_k-\beta_0\rho_k\right)}.
\eeq
where $v_{\zeta_i}=\rho_{\zeta_i}^{-1}$, $m_{z_j}=m_k$ and $v_{z_j}=\rho_{k}^{-1}$ if $z_j=k$, $\Vb_{\zb}=\diag{\vb_{\zb}}$, and $\Vb_{\zeta}=\diag{\vb_{\zeta}}$. The evidence $p(\gb|\Mc)$ does not depend on the unknowns.

In \cite{chapdelaine20163DBayesian}, an algorithm is proposed to compute the Maximum-A-Posteriori (MAP) estimator for this posterior distribution in 3D X-ray CT. Another possible estimator in decision theory is Minimum Mean-Square Error (MMSE), which is Posterior Mean (PM). The calculation of PM can be achieved by MCMC methods which generate samples of the posterior distribution \eqref{eq:posterior distribution}, typically using a Gibbs sampler \cite{ayasso2010bayesian, zhao2016joint}. The problem is that the computational complexity of these methods is unaffordable in large 3D inverse problems such as 3D X-ray CT \cite{zhao2016joint, chapdelaine20163DBayesian}. Variational Bayesian approach (VBA) enables to alleviate the cost for PM calculation, by computing an analytical approximation of the true posterior distribution \eqref{eq:posterior distribution} \cite{SQ06, pereyra2016survey}. This approximation is chosen sufficiently simple to compute PM. For instance, it can be a fully-factorized density, which corresponds to mean-field approximation (MFA) \cite{pereyra2016survey}. VBA is known to well-estimate PM by computing an approximate posterior distribution \cite{giordano2015linear, pereyra2016survey, pereyra2017maximum}. Once its factorization is chosen, the approximate posterior distribution $q$ has to minimize Kullback-Leibler divergence \cite{SQ06, pereyra2016survey} : 
\beq
KL(q(\psib) || p(\psib|\gb))=\int_{\psib} q(\psib)\ln\left(\frac{q(\psib)}{p(\psib|\gb)}\right)d\psib=\ln\left(p(\gb|\Mc)\right)-\Fc(q(\psib)) 
\label{eq:Kullback-Leibler divergence}
\eeq
where
\beq
\barr{ll}
\Fc(q(\psib))&=\int_{\psib} q(\psib) \ln\left(\frac{p(\gb,\psib)}{q(\psib)}\right) d\psib \\
&=-\int_{\psib} q(\psib)\ln\left(q(\psib)\right)d\psib+\int_{\psib} q(\psib) \ln\left(p(\gb,\psib)\right) d\psib
\earr
\label{eq: Energie libre négative et entropie}
\eeq
is free negative energy \cite{SQ06,AyDj10}. The entropy of approximate posterior distribution $q$ is defined as
\beq
\Hc(q(\psib))=-\int_{\psib} q(\psib)\ln\left(q(\psib)\right)d\psib.
\eeq
In the next section, we present an algorithm implementing VBA for computing PM with Gauss-Markov-Potts prior model.

\section{Variational Bayesian Approach with Gauss-Markov-Potts prior}
\label{sec:Application de l'approche bayésienne variationnelle en tomographie par rayons X pour la détermination des incertitudes}

For
\[
\psib=(\fb,\rhob_{\zeta},\zb,\mb,\rhob),
\]
we choose an approximate posterior distribution of the form
\begin{align}
q(\fb,\rhob_{\zeta},\zb, \mb, \rhob)
&=\prod_{j=1}^{N} q_{f_j}(f_{j}|z_j) \times
\prod_{j=1}^{N} q_{z_j}(z_{j}) \nonumber \\
&\times\prod_{i=1}^{M} q_{\rho_{\zeta_{i}}}(\rho_{\zeta_{i}})\times
\prod_{k=1}^{K} q_{m_{k}}(m_k) 
\times
\prod_{k=1}^{K} q_{\rho_{k}}(\rho_k).
\label{eq:Loi approchante des inconnues avec le modèle direct usuel}
\end{align}
This approximation performs a partial separation, since the dependence between $f_j$ and $z_j$ is preserved \cite{AyDj10}. In our experiments, we have noticed an unsteady behaviour if this dependence is broken. This is because separating $f_j$ and $z_j$ leads to a too gross approximation \cite{pereyra2016survey}.

Minimizing Kullback-Leibler divergence with respect to each factor leads to
\beq
\left\{
\barr{ll}
q_{f_j}(f_{j}|z_j=k)=\Nc(f_{j}|\tilde{m}_{jk}, \tilde{v}_{jk}), \forall k \\
& \\
q_{z_j}(z_j)
=\frac{\expf{\sum_{k=1}^{K}
\left(\tilde{\alpha}_{jk}+\gamma_{0}\sum_{i \in \Vc(j)}q_{z_{i}}(k)\right)\delta(z_j-k)}}{\sum_{k=1}^{K}\expf{
\tilde{\alpha}_{jk}+\gamma_{0}\sum_{i \in \Vc(j)}q_{z_{i}}(k)}} \\
&\\
q_{\rho_{\zeta_{i}}}(\rho_{\zeta_{i}})=\Gc(\rho_{\zeta_{i}}|\tilde{\alpha}_{\zeta_{0_{i}}}, \tilde{\beta}_{\zeta_{0_{i}}}) \\
& \\
q_{m_{k}}(m_k)=\Nc(m_k|\tilde{m}_{0_{k}}, \tilde{v}_{0_{k}}) \\
& \\
q_{\rho_{k}}(\rho_k)=\Gc(\rho_k|\tilde{\alpha}_{0_{k}}, \tilde{\beta}_{0_{k}}) 
\earr
\right.
\label{eq:lois approchantes avec le modèle MGI et le modèle direct usuel}
\eeq
where $q_{z_{i}}(k)$ in the expression of $q_{z_j}(k)$ is the value of $q_{z_{i}}(k)$ at previous iteration. The algorithm in figure \ref{fig:Iterative algorithm to compute approximating distribution : order} turns into an iterative updating of the parameters of the distributions in equation \eqref{eq:lois approchantes avec le modèle MGI et le modèle direct usuel}. The updating formulae are given hereafter.

We introduce \textit{digamma} function
\beq
\psi(x)=\frac{\Gamma'(x)}{\Gamma(x)},
\label{eq:fonction digamma}
\eeq
as the expectation of number $N_k$ of voxels in class $k$ with respect to approximate distribution $q$
\beq
\mathbb{E}_{q_{\zb}}(N_{k}(\Zb))=\sum_{j=1}^{N}q_{z_j}(k),
\label{eq:espérance par rapport à la loi approchante du nombre de voxels dans la classe k}
\eeq
and several auxiliary variables :
\beq
\left\{
\barr{ll}
\tilde{m}_j=\sum_{k=1}^{K}\tilde{m}_{jk}q_{z_j}(k) \\
\\
\tilde{v}_j=\sum_{k=1}^{K}\tilde{v}_{jk}q_{z_j}(k) \\
\\
\tilde{m}_{j}^{(2)}=\sum_{k=1}^{K}\left(\tilde{m}_{jk}-\tilde{m}_{j}\right)^{2}q_{z_j}(k)=\sum_{k=1}^{K}\tilde{m}_{jk}^{2}q_{z_j}(k)-\tilde{m}_{j}^{2} \\
\\
\tilde{v}_{j}^{(2)}=\tilde{v}_j+\tilde{m}_{j}^{(2)}
\\
\tilde{v}_{\zeta_i}=\frac{\tilde{\beta}_{\zeta_{0_{i}}}}{\tilde{\alpha}_{\zeta_{0_{i}}}}
\earr
\right.
\label{eq:variables intermédiaires pour l'approche bayésienne variationnelle, avec le modèle direct usuel}
\eeq
and $\tilde{\Vb}_{\zeta}=\diag{\tilde{\vb}_{\zeta}}$. 

After calculations, the entropy of approximate distribution $q$ reads
\begin{align}
&\Hc(q(\fb, \rhob_{\zeta}, \zb, \mb, \rhob))=\frac{N}{2}(1+\ln(2\pi))+\frac{1}{2}\sum_{j=1}^{N}\sum_{k=1}^{K}\ln\left(\tilde{v}_{jk}\right)q_{z_j}(k)-\sum_{j=1}^{N}\sum_{k=1}^{K}q_{z_j}(k)\ln\left(q_{z_j}(k)\right) \nonumber \\
&+\sum_{i=1}^{M}\left[\ln(\Gamma(\tilde{\alpha}_{\zeta_{0_{i}}}))-\ln(\tilde{\beta}_{\zeta_{0_{i}}})+\tilde{\alpha}_{\zeta_{0_{i}}}-\left(\tilde{\alpha}_{\zeta_{0_{i}}}-1\right)\psi\left(\tilde{\alpha}_{\zeta_{0_{i}}}\right)\right] \nonumber \\
&+\frac{K}{2}(1+\ln(2\pi))+\frac{1}{2}\sum_{k=1}^{K}\ln\left(\tilde{v}_{0_{k}}\right)+\sum_{k=1}^{K}\left[\ln(\Gamma(\tilde{\alpha}_{0_k}))-\ln(\tilde{\beta}_{0_k})+\tilde{\alpha}_{0_k}-\left(\tilde{\alpha}_{0_k}-1\right)\psi\left(\tilde{\alpha}_{0_k}\right)\right]
\label{eq:entropie de la loi approchante avec le modèle direct usuel}
\end{align}
and the expectation of the joint distribution of the data and the unknowns, with respect to approximate distribution $q$, is
\begin{align}
&\mathbb{E}_{q}(\ln\left((p(\gb;\fb, \rhob_{\zeta}, \zb, \mb, \rhob|\Mc))\right)=-\frac{M}{2}\ln(2\pi)-\frac{1}{2}\sum_{i=1}^{M}\left[\ln(\tilde{\beta}_{\zeta_{0_{i}}})-\psi\left(\tilde{\alpha}_{\zeta_{0_{i}}}\right)\right] \nonumber \\
&-\frac{1}{2}\|\gb-\Hb\tilde{\mb}\|_{\tilde{\Vb}_{\zeta}}^{2}-\frac{1}{2}\sum_{j=1}^{N}\tilde{v}_{j}^{(2)}\left[\Hb^{T}\tilde{\Vb}_{\zeta}^{-1}\Hb\right]_{jj} \nonumber \\
&-\frac{N}{2}\ln(2\pi)-\frac{1}{2}\sum_{j=1}^{N}\sum_{k=1}^{K}\left(\frac{\tilde{\alpha}_{0_{k}}}{\tilde{\beta}_{0_{k}}}
\left[\tilde{v}_{jk}+\tilde{v}_{0_{k}}+\left(\tilde{m}_{jk}-\tilde{m}_{0_{k}}\right)^{2}\right]
+\ln(\tilde{\beta}_{0_{k}})-\psi(\tilde{\alpha}_{0_{k}})\right)q_{z_j}(k) \nonumber \\
&-\ln\left(Z(\alphab, \gamma_0)\right)+\sum_{j=1}^{N}\sum_{k=1}^{K}\left(\alpha_k+\gamma_0\sum_{i \in \Vc(j)}q_{z_{i}}(k)\right)q_{z_j}(k) \nonumber \\
&-M\left(\ln(\Gamma(\alpha_{\zeta_0}))-\alpha_{\zeta_0}\ln(\beta_{\zeta_0})\right)-(\alpha_{\zeta_0}-1)\sum_{i=1}^{M}(\ln(\tilde{\beta}_{\zeta_{0_{i}}})-
\psi(\tilde{\alpha}_{\zeta_{0_{i}}}))-\beta_{\zeta_0}\sum_{i=1}^{M}\frac{\tilde{\alpha}_{\zeta_{0_{i}}}}{\tilde{\beta}_{\zeta_{0_{i}}}} \nonumber \\
&-\frac{K}{2}\ln(2\pi v_0)-\frac{1}{2v_0}\sum_{k=1}^{K}\left(\tilde{v}_{0_{k}}+\left(\tilde{m}_{0_{k}}-m_0\right)^{2}\right) \nonumber \\
&-K\left(\ln(\Gamma(\alpha_0))-\alpha_0\ln(\beta_0)\right)-(\alpha_0-1)\sum_{k=1}^{K}(\ln(\tilde{\beta}_{0_{k}})-
\psi(\tilde{\alpha}_{0_{k}}))-\beta_0\sum_{k=1}^{K}\frac{\tilde{\alpha}_{0_{k}}}{\tilde{\beta}_{0_{k}}}.
\label{eq:espérance de la loi jointe des données et des inconnues par rapport à la loi approchante avec le modèle direct usuel}
\end{align}
The stopping criterion of the algorithm in figure \ref{fig:Iterative algorithm to compute approximating distribution : order} is free negative energy :
\beq
\Fc(q(\fb, \rhob_{\zeta}, \zb, \mb, \rhob))=\Hc(q(\fb, \rhob_{\zeta}, \zb, \mb, \rhob))+\mathbb{E}_{q}(\ln\left((p(\gb;\fb, \rhob_{\zeta}, \zb, \mb, \rhob|\Mc))\right)
\label{eq:énergie libre négative avec Gauss-Markov-Potts}
\eeq
from which constants are removed. At the end of the algorithm, the unknowns are estimated by their expectation according to the approximate distribution, excepted for the labels which are estimated by maximum a posteriori, due to the fact that they are discrete variables :
\beq
\left\{
\barr{ll}
\hat{z}_j=\arg\max_{k}{\left\{q_{z_j}(k)\right\}} \\
\hat{f}_j=\tilde{m}_{jk} \mbox{~~avec~~} k=\hat{z}_j \\
\hat{\rho}_{\zeta_i}=\frac{\tilde{\alpha}_{\zeta_{0_{i}}}}{\tilde{\beta}_{\zeta_{0_{i}}}} \\
\hat{m}_k=\tilde{m}_{0_k} \\
\hat{\rho}_k=\frac{\tilde{\alpha}_{0_{k}}}{\tilde{\beta}_{0_{k}}}
\earr
\right.
\label{eq:estimations à la fin du VBA}
\eeq

In the algorithm, the updating order of the parameters of approximate distributions is important. This order is shown in figure \ref{fig:Iterative algorithm to compute approximating distribution : order}. The distributions of the variables which are approximated as independent are immediatly replaced by their updates. On the opposite, the updating of joint approximate distribution of the volume and the labels
\beq
q_{\fb,\zb}^{(t)}(\fb,\zb)=q_{\fb}^{(t)}(\fb|\zb)q_{\zb}^{(t)}(\zb)
\eeq
involves two steps : the update of $q_{\fb}$ and the one of $q_{\zb}$. For this reason, $q_{\zb}^{(t)}(\zb)$ is updated using $q_{\fb}^{(t-1)}(\fb|\zb)$.

\bfig[htb]
\bcc
\begin{picture}(400,350)

  \put(20,290){\framebox(380,30){compute $q_{\fb}^{(t)}(\fb|\zb)$ according to $q_{\fb}^{(t-1)}(\fb|\zb),q_{\zb}^{(t-1)}(\zb),q_{\rhob_{\zeta}}^{(t-1)}(\rhob_{\zeta}),q_{\mb}^{(t-1)}(\mb),q_{\rhob}^{(t-1)}(\rhob)$}}
  \put(195,290){\vector(0,-1){20}}
  \put(20,240){\framebox(380,30){compute $q_{\zb}^{(t)}(\zb)$ according to $q_{\fb}^{(t-1)}(\fb|\zb),q_{\zb}^{(t-1)}(\zb),q_{\rhob_{\zeta}}^{(t-1)}(\rhob_{\zeta}),q_{\mb}^{(t-1)}(\mb),q_{\rhob}^{(t-1)}(\rhob)$}}
  \put(195,240){\vector(0,-1){20}}
  \put(20,190){\framebox(380,30){compute $q_{\rhob_{\zeta}}^{(t)}(\rhob_{\zeta})$ according to $q_{\fb}^{(t)}(\fb|\zb),q_{\zb}^{(t)}(\zb),q_{\mb}^{(t-1)}(\mb),q_{\rhob}^{(t-1)}(\rhob)$}}
  \put(195,190){\vector(0,-1){20}}
  \put(20,140){\framebox(380,30){compute $q_{\mb}^{(t)}(\mb)$ according to $q_{\fb}^{(t)}(\fb|\zb),q_{\zb}^{(t)}(\zb),q_{\rhob_{\zeta}}^{(t)}(\rhob_{\zeta}),q_{\rhob}^{(t-1)}(\rhob)$}}
  \put(195,140){\vector(0,-1){20}}
  \put(20,90){\framebox(380,30){compute $q_{\rhob}^{(t)}(\rhob)$ according to $q_{\fb}^{(t)}(\fb|\zb),q_{\zb}^{(t)}(\zb),q_{\rhob_{\zeta}}^{(t)}(\rhob_{\zeta}),q_{\mb}^{(t)}(\mb)$}}
  \put(195,90){\vector(0,-1){20}}
  \put(20,40){\makebox(350,30){$t:=t+1$}}
  \put(195,40){\line(0,-1){20}}
  \put(195,20){\line(-1,0){200}}
  \put(-5,20){\line(0,1){320}}
  \put(-5,340){\line(1,0){200}}
  \put(195,340){\vector(0,-1){20}}
   
\end{picture}
\ecc
\caption{Iterative algorithm to compute approximating distribution $q(\fb,\rhob_{\zeta},\zb,\mb, \rhob)$}
\label{fig:Iterative algorithm to compute approximating distribution : order}
\efig

\subsection{Approximate distribution for the volume}
\label{sec:Loi approchante pour le volume}

By minimizing Kullback-Leibler divergence with respect to $q_{f_j}$, we have :
\beq
q_{f_j}(f_{j}|z_j=k)=\Nc(f_{j}|\tilde{m}_{jk}, \tilde{v}_{jk}), \forall j, \forall k
\eeq
where
\beq
\left\{
\barr{ll}
\tilde{v}_{jk}=\left(\frac{\tilde{\alpha}_{0_{k}}}{\tilde{\beta}_{0_{k}}}+\left[\Hb^{T}\tilde{\Vb}_{\zeta}^{-1}\Hb\right]_{jj}\right)^{-1} \\
\tilde{m}_{jk}=\tilde{m}_j+\tilde{v}_{jk}\left(\frac{\tilde{\alpha}_{0_{k}}}{\tilde{\beta}_{0_{k}}}(\tilde{m}_{0_{k}}-\tilde{m}_j)+\left[\Hb^{T}\tilde{\Vb}_{\zeta}^{-1}(\gb-\Hb\tilde{\mb})\right]_{j}\right)
\earr
\right.
\label{eq:paramètres de la loi approchante du volume avec le modèle direct usuel}
\eeq 

\subsection{Approximate distribution for the labels}
\label{sec:Loi approchante pour les labels}

By minimizing Kullback-Leibler divergence with respect to $q_{z_j}$, we have :
\beq
q_{z_j}(z_j)
=\frac{\expf{\sum_{k=1}^{K}
\left(\tilde{\alpha}_{jk}+\gamma_{0}\sum_{i \in \Vc(j)}q_{z_{i}}(k)\right)\delta(z_j-k)}}{\sum_{k=1}^{K}\expf{
\tilde{\alpha}_{jk}+\gamma_{0}\sum_{i \in \Vc(j)}q_{z_{i}}(k)}}
\eeq
where $q_{z_{i}}(k), \forall i \in \Vc(j),$ is the value of $q_{z_{i}}(k)$ at the previous iteration of the algorithm presented in figure \ref{fig:Iterative algorithm to compute approximating distribution : order}. We have
\begin{align}
\tilde{\alpha}_{jk}&=\alpha_{k}-\frac{1}{2}\left(\frac{\tilde{\alpha}_{0_{k}}}{\tilde{\beta}_{0_{k}}}
\left[\tilde{v}_{jk}+\tilde{v}_{0_{k}}+
\left(\tilde{m}_{jk}-\tilde{m}_{0_{k}}\right)^{2}\right]
+\ln(\tilde{\beta}_{0_{k}})-
\psi(\tilde{\alpha}_{0_{k}})\right) \nonumber \\
&-\frac{1}{2}\left(\left(\tilde{v}_{jk}+\tilde{m}_{jk}^{2}\right)\left[\Hb^{T}\tilde{\Vb}_{\zeta}^{-1}\Hb\right]_{jj}-2\tilde{m}_{jk}\left(\tilde{m}_{j}\left[\Hb^{T}\tilde{\Vb}_{\zeta}^{-1}\Hb\right]_{jj}+\left[\Hb^{T}\tilde{\Vb}_{\zeta}^{-1}(\gb-\Hb\tilde{\mb})\right]_{j}\right)\right) \nonumber \\
&+\frac{1}{2}\ln\left(\tilde{v}_{jk}\right)
\label{eq:paramètres de la loi approchante des labels pour MGI}
\end{align}
It is worth to notice that this step does not imply the calculation of diagonal coefficients $\left[\Hb^{T}\tilde{\Vb}_{\zeta}^{-1}\Hb\right]_{jj}, \forall j,$ and of the backprojection of the errors, since they have been computed before to update the approximate distribution of the volume (see the algorithm shown in figure \ref{fig:Iterative algorithm to compute approximating distribution : order}).

\subsection{Approximate distribution of inverse variances of the uncertainties}
\label{sec:Loi approchante pour les variances des incertitudes}

By minimizing Kullback-Leibler divergence with respect to $q_{\rho_{\zeta_{i}}}$, we have : 
\beq
q_{\rho_{\zeta_{i}}}(\rho_{\zeta_{i}})=\Gc(\rho_{\zeta_{i}}|\tilde{\alpha}_{\zeta_{0_{i}}}, \tilde{\beta}_{\zeta_{0_{i}}})
\eeq 
where paramereters are updated by formulae : 
\beq
\left\{
\barr{ll}
\tilde{\alpha}_{\zeta_{0_{i}}}=\alpha_{\zeta_0}+\frac{1}{2} \\
\tilde{\beta}_{\zeta_{0_{i}}}=\beta_{\zeta_0}+\frac{1}{2}
\left(\left(g_{i}-\left[\Hb\tilde{\mb}\right]_{i}\right)^{2}+
\left(\Hb\tilde{\Vb}^{(2)}\Hb^{T}\right)_{ii}\right)
\earr
\right.
, \forall i \in \lbrace 1, \dots, M \rbrace.
\label{eq:paramètres de la loi approchante pour les variances des incertitudes globales}
\eeq
and where $\tilde{\Vb}^{(2)}=\diag{\tilde{\vb}^{(2)}}$.

\subsection{Approximate distribution for the means of the classes}
\label{sec:Loi approchante pour les moyennes des classes}

By minimizing Kullback-Leibler divergence with respect to $q_{m_k}$, we have :  
\beq
q_{m_k}(m_k)=\Nc(m_k|\tilde{m}_{0_{k}}, \tilde{v}_{0_{k}})
\eeq 
where
\beq
\left\{
\barr{ll}
\tilde{v}_{0_{k}}=\left(\frac{1}{v_0}+
\frac{\tilde{\alpha}_{0_{k}}}{\tilde{\beta}_{0_{k}}}\sum_{j=1}^{N}q_{z_j}(k)\right)^{-1} \\
\tilde{m}_{0_{k}}=\tilde{v}_{0_{k}}\left(\frac{m_0}{v_0}+\frac{\tilde{\alpha}_{0_{k}}}{\tilde{\beta}_{0_{k}}}\sum_{j=1}^{N}\tilde{m}_{jk}q_{z_j}(k)\right)
\earr
\right.
\label{eq:paramètres de la loi approchante des moyennes des classes pour le modèle GMP-MGI}
\eeq

\subsection{Approximate distribution for the inverses of variances of the classes}
\label{sec:Loi approchante pour les variances des classes}

By minimizing Kullback-Leibler divergence with respect to $q_{\rho_k}$, we have : 
\beq
q_{\rho_k}(\rho_k)=\Gc(\rho_k|\tilde{\alpha}_{0_{k}}, \tilde{\beta}_{0_{k}}) 
\eeq 
where
\beq
\left\{
\barr{ll}
\tilde{\alpha}_{0_{k}}=\alpha_{0}+\frac{1}{2}\sum_{j=1}^{N}q_{z_j}(k)=\alpha_{0}+\frac{1}{2}\mathbb{E}_{q_{\zb}}(N_{k}(\Zb)) \\
\tilde{\beta}_{0_{k}}=\beta_{0}+\frac{1}{2}\sum_{j=1}^{N}\left(\tilde{v}_{0_{k}}+\tilde{v}_{jk}+\left(\tilde{m}_{jk}-\tilde{m}_{0_{k}}\right)^{2}\right)q_{z_j}(k)
\earr
\right.
\label{eq:paramètres de la loi approchante des variances des classes pour le modèle GMP-MGI}
\eeq

\subsection{Fixing the parameters for Variational Bayesian Approach}
\label{sec:Réglage des paramètres}

The strategies to fix the parameters for VBA are different from JMAP, since a different estimator is computed. This is particularly the case for parameters $(\alpha_{\zeta_0},\beta_{\zeta_0},\alpha_0,\beta_0)$.

Through our experiments, Jeffreys' priors \cite{jeffreys1946invariant}, which is non-informative, for $\rhob_{\zeta}$ and $\rhob$ gives the best results \cite{AyDj10}.  Gamma distribution $\Gc(\rho|\alpha,\beta)$ is non-informative fixing $\alpha=0$ et $\beta=0$ but leads to improper prior. In order to keep our priors proper, $(\alpha_{\zeta_0},\beta_{\zeta_0},\alpha_0,\beta_0)$ are fixed near zero : $\alpha_{\zeta_0}\ll 1, \beta_{\zeta_0}\ll 1, \alpha_0\ll 1$ and $\beta_0\ll 1$. To take the SNR : \textit{Signal-to-Noise Ratio} into account, we comply with the following relation
\beq
\mathbb{E}(\rho_{\zeta_i}|\alpha_{\zeta_0},\beta_{\zeta_0})=\frac{\alpha_{\zeta_0}}{\beta_{\zeta_0}}=\mathbb{E}(\rho_k|\alpha_0,\beta_0)\times 10^{\frac{SNR}{10}}=\frac{\alpha_0}{\beta_0}\times 10^{\frac{SNR}{10}}.
\eeq
Moreover, in order to avoid "NaN"-values in our computations, $(\alpha_{\zeta_0},\beta_{\zeta_0},\alpha_0,\beta_0)$ are fixed such that
\beq
\frac{\alpha_{\zeta_0}}{\beta_{\zeta_0}}\leq 1 \mbox{~~et~~} \frac{\alpha_0}{\beta_0}\leq 1.
\eeq
Other parameters are fixed as in JMAP \cite{chapdelaine20163DBayesian}. From initial volume $\fb^{(0)}$, $m_0$ is fixed by
\beq
m_0=\frac{1}{2}\left(\min_{j}{f_{j}^{(0)}}+\max_{j}{f_{j}^{(0)}}\right),
\eeq
$v_0$ is fixed sufficiently large such that $m_k, \forall k$, can take all possible values in the set of gray levels of the volume, $\gamma_0$ is sufficiently large to promote compact classes, and, from initial segmentation $\zb^{(0)}$ of $\fb^{(0)}$, we fix 
\beq
\alpha_k=\ln\left(\frac{N_{k}^{(0)}}{N}\right)
\eeq 
where $N_{k}^{(0)}$ is the number of voxels in class $k$ in initial segmentation $\zb^{(0)}$. Like in \cite{chapdelaine20163DBayesian}, the number of classes $K$ is supposed to be known a priori.

\subsection{Initialization}
\label{sec:Initialisation}

The initialization of the algorithm is crucial to ensure its convergence. Based on an initial reconstruction $\fb^{(0)}$ obtained for instance by filtered backprojection \cite{feldkamp1984practical}, an initial segmentation $\zb^{(0)}$ is performed by applying a fast method \cite{macqueen1967some, otsu1979thresholds, kurugollu2001color}. From this initial segmentation, initial means $\mb^{(0)}$ and variances $\vb^{(0)}$ of the classes are computed. We initialize means $\tilde{m}_{jk}, \forall j,k$ by $f_{j}^{(0)}$ if voxel $j$ is in class $k$ at initialization, and $m_{k}^{(0)}$ otherwise :
\beq
\tilde{m}_{jk}^{(0)}=
\left\{
\barr{ll}
f_{j}^{(0)} \mbox{~~if~~} z_{j}^{(0)}=k \\
m_{k}^{(0)} \mbox{~~otherwise~~}
\earr
\right.
\eeq
Probabilities to be in each class are initialized by $0$ or $1$ :
\beq
q_{z_j}(k)^{(0)}=
\left\{
\barr{ll}
1 \mbox{~~if~~} z_{j}^{(0)}=k \\
0 \mbox{~~otherwise~~}
\earr
\right.
\eeq
Inspired by their updating formula \eqref{eq:paramètres de la loi approchante du volume avec le modèle direct usuel}, we initialize variances $\tilde{v}_{jk}$ by :
\beq
\tilde{v}_{jk}^{(0)}=\left(\frac{1}{v_{k}^{(0)}}+\frac{\alpha_{\zeta_0}}{\beta_{\zeta_0}}\left[\Hb^T\Hb\right]_{jj}\right)^{-1}
\eeq
The approximate distributions of $\rhob_{\zeta}$, $\mb$ and $\rhob$
are initialized by their conditional distribution given the other unknowns. These expressions are given in \cite{chapdelaine20163DBayesian} :
\beq
\tilde{\alpha}_{\zeta_{0_i}}^{(0)}=\alpha_{\zeta_0}+\frac{1}{2}, \forall i,
\eeq
\beq
\tilde{\beta}_{\zeta_{0_i}}^{(0)}=\beta_{\zeta_0}+\frac{1}{2}\left(g_i-\left[\Hb\fb^{(0)}\right]_i\right)^2, \forall i,
\eeq
\beq
\tilde{v}_{0_k}^{(0)}=\left(\frac{1}{v_0}+\frac{N_{k}^{(0)}}{v_{k}^{(0)}}\right)^{-1}, \forall k,
\eeq
\beq
\tilde{m}_{0_k}^{(0)}=\tilde{v}_{0_k}^{(0)}\left(\frac{m_0}{v_0}+N_{k}^{(0)}\frac{m_{k}^{(0)}}{v_{k}^{(0)}}\right), \forall k,
\eeq
\beq
\tilde{\alpha}_{0_k}^{(0)}=\alpha_0+\frac{N_{k}^{(0)}}{2}, \forall k,
\eeq
\beq
\tilde{\beta}_{0_k}^{(0)}=\beta_0+\frac{N_{k}^{(0)}}{2}v_{k}^{(0)}, \forall k.
\eeq

\section{Conclusion and perspectives}
\label{sec:Conclusion et perspectives}

In this paper, we have presented a full algorithm for joint estimation of reconstruction and uncertainties in linear inverse problems regularized by Gauss-Markov-Potts prior model. Perspectives for this work are applications to large 3D inverse problems such as 3D X-ray CT, for which implementation problems will be discussed in future works.

\bibliographystyle{ieeetr}
\bibliography{ChapdelaineBibliographie}

\end{document}